\documentclass[11pt,reprint,aps,pre]{revtex4-2}
\usepackage{graphicx}
\usepackage{amsmath,amssymb,bm}
\usepackage{hyperref}
\usepackage{soul,color}
\usepackage{siunitx}
\usepackage[normalem]{ulem}
\usepackage[capitalize]{cleveref}

\newcommand{\nn}{\nonumber\\}
\newcommand*\diff{\mathop{}\!\mathrm{d}}

\sethlcolor{yellow}

\begin{document}

\title{Charging dynamics of electric double layer capacitors including beyond-mean-field electrostatic correlations}

\author{David Fertig}
\email{david.fertig@nmbu.no}
\affiliation{Institute of Physics, Norwegian University of Life Sciences, \AA s, Norway}
 
\author{Mathijs Janssen}
\email{mathijs.a.janssen@nmbu.no}
\affiliation{Institute of Physics, Norwegian University of Life Sciences, \AA s, Norway}

\date{\today}

\begin{abstract}
Electric double layer (EDL) formation underlies the functioning of supercapacitors and several other electrochemical technologies. 
Here, we study how the EDL formation near two flat blocking electrodes separated by $2L$ is affected by beyond-mean-field Coulombic interactions, which can be substantial for electrolytes of high salt concentration or with multivalent ions.
Our model combines the Nernst-Planck and Bazant-Storey-Kornyshev (BSK) equations; the latter is a modified Poisson equation with a correlation length $\ell_c$.
In response to a voltage step, the system charges exponentially with a characteristic timescale $\tau$ that depends nonmonotonically on $\ell_c$. 
For small $\ell_c$, $\tau$ is given by the BSK capacitance times a dilute electrolyte's resistance, in line with [Zhao, Phys. Rev. E {\bf 84} (2011)]; here, $\tau$ decreases with increasing $\ell_c$. 
Increasing the correlation length beyond $\ell_c\approx L^{2/3}\lambda_D^{1/3}$, with $\lambda_D$ the Debye length, $\tau$ reaches a minimum, rises as $\tau\propto \lambda_D\ell_c/D$, and plateaus at $\tau=4L^2/(\pi^2 D)$.
Our results imply that strongly correlated, strongly confined electrolytes---ionic liquids in the surface force balance apparatus, say---move slower than predicted so far.
\end{abstract}

\maketitle

\section{Introduction}
In their seminal 1923 work \cite{debye_pz1_1923}, Debye and H\"{u}ckel showed how Coulombic interactions among ions lead to correlations in their positions.
A cation in a bulk electrolyte is surrounded by an anionic cloud, the size of which being set by a concentration-dependent length now called the Debye length, $\lambda_D$. 
Effects of bulk electrostatic correlations are widely observed, from colloid science and biology to plasma physics \cite{levin2002electrostatic}.

In electrochemistry, even more important than an electrolyte's static bulk properties are its dynamics near electrodes \cite{santos_ea_2024}.
Debye and H\"{u}ckel showed that 
correlations reduce a dilute electrolyte's bulk conductance to $\Lambda=\Lambda_0-K\sqrt{I}$ \cite{debye_pz_1923}, with $\Lambda_0$ the conductance at infinite dilution, $I$ the ionic strength, and $K$ a constant determined by Onsager~\cite{onsager_pz_1926,onsager_pz_1927}.
Next, Debye and Falkenhagen showed that, for harmonic applied electric fields, the conductance decreases with increasing driving frequency.
Debye and his contemporaries modeled the electrostatic potential around a bulk ion by the Poisson-Boltzmann (PB) and Poisson-Nernst-Planck (PNP) equations \cite{debye_pz1_1923,debye_pz_1923,onsager_pz_1926,onsager_pz_1927,debye_zeapc_1928}.
As this potential is small compared to the thermal voltage ($\approx\SI{25}{\milli\volt}$ at room temperature), they could linearize the PB and PNP equations to, in modern parlance, the Debye-H\"{u}ckel and Debye-Falkenhagen (DF) equations.
A century later, correlated electrolytes can be modeled in far greater detail with molecular simulations and statistical mechanics.
These methods, however, cannot be used to simulate electrolyte dynamics in complicated or large geometries and they do not yield analytical expressions for key observables. 
The PB and PNP equations and their linearizations are therefore still used today to gain physical insight.
Moreover, these equations form the starting point for several extensions discussed below.

\begin{figure}[b]
    \centering
	\includegraphics[width =\linewidth]{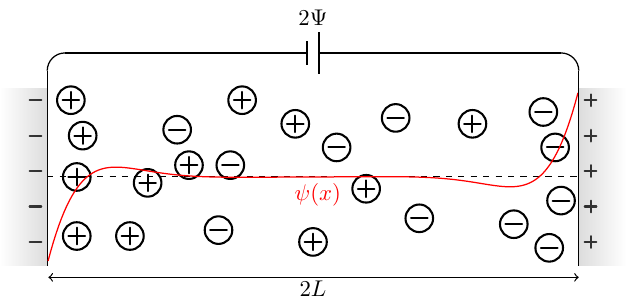}
    \caption{The canonical setup (not to scale) for studying EDL formation near electrodes: two blocking flat parallel electrodes with a binary electrolyte in between. The curve shows the potential $\psi(x)$ for $\ell_c/\lambda_D=4$ and $L/\lambda_D=20$, at which the EDL overscreens the electrode charge.}\label{fig:setup}
\end{figure}

The canonical setup to study interfacial electrolyte dynamics contains an electrolyte between two flat parallel electrodes separated by $2L$, subject to a time-dependent potential difference $2\Psi$, see \cref{fig:setup}.
Pioneering work by Macdonald~\cite{macdonald_pr_1953} considered this setup subject to a small harmonic potential.
Solving the DF equation, he found that the admittance contains a characteristic timescale $\lambda_D L/D$, with $D$ the ionic diffusivity, which is the $RC$ time of the system.
Decades later, Bazant, Thornton, and Ajdari showed that the same timescale sets the late-time relaxation of the canonical setup in response to a small step potential difference \cite{bazant_pre_2004}.
The full transient response contains infinitely many exponentially decaying modes, whose timescales are all proportional to the Debye time $\lambda_D^2/D$ \cite{janssen_pre_2018,palaia2019charged}.
Molecular simulations \cite{asta_jcp_2019,scalfi_arpc_2021,ahrensivers_jcp_2022,pireddu_prl_2023} largely confirmed the DF analyses of Refs.~\cite{macdonald_pr_1953,bazant_pre_2004,janssen_pre_2018,palaia2019charged} in the relevant parameter regime of small applied potentials, dilute electrolytes, and  monovalent ions.
However, experiments of the transient \cite{beunis_apl_2007,zhao_jpcc_2024}, harmonic \cite{kortschot_jpcc_2014}, and cyclic voltammetric \cite{nakamura2014structural,ojha2020double} response of the canonical setup contain unexplained effects, challenging simulators and theorists to extend their models.

For higher salt concentrations or for multivalent ions, electrostatic correlations affect an electrolyte's structure near electrodes. 
Statistical mechanics and molecular simulations show that the first layer of counterions near a charged electrode can contain more ionic charge than the electrode carries electrons (or holes). 
Such ``overscreening'' leads to oscillating potential profiles (see the red line in \cref{fig:setup}) and capacitance decrease that cannot be captured by the PB equation \cite{fertig_pccp_2020}.
Extensions were thus developed to better understand how electrostatic correlations affect an electrolyte's structure and dynamics near electrodes
\cite{bazant_pre_2011,storey_pre_2012,Hatlo_2012,balu2018role,desouza_jpcc_2020,souza2020interfacial,gupta_jpcc_2020,gupta_prl_2020}.
Most notably, Bazant, Storey, and Kornyshev (BSK) proposed a correction to the linear response between electric field and polarization,
yielding a modified Poisson equation  [\emph{viz}.~\cref{Eq:mP}] now called the BSK equation \cite{bazant_pre_2011,storey_pre_2012}.
The BSK equation contains a correlation length $\ell_c$ reflecting a dense electrolyte's short-range electrostatic correlations \cite{desouza_jpcc_2020}.
 {In Ref.}~\cite{bazant_pre_2011}, { the correlation length $\ell_c$ is in the order of the ion size. Conversely, de Souza and Bazant} \cite{desouza_jpcc_2020}  {fitted their model to MC simulations and found that $\ell_c$ depended on electrolyte concentration but not the ion size [see Eqs. (13)-(15) therein].}
The BSK equation predicts oscillating potential profiles for $\ell_c>\lambda_D$ and, for small applied potentials, a correlation-induced correction to the EDL capacitance, reading
\begin{equation}\label{eq:BSKcapacitance}
    C=C_{DH}\dfrac{\sqrt{1+2\ell_c/\lambda_D}}{1+\ell_c/\lambda_D},
\end{equation}
where $\varepsilon$ is the permittivity and $C_{DH}=\varepsilon/\lambda_D$ is the Debye-H\"{u}ckel capacitance per unit electrode area of an uncorrelated electrolyte.

To study the effect of correlations on EDL dynamics in the canonical setup, one could couple the BSK and Nernst-Planck (NP) [\emph{viz}.~\cref{eq:nernstplanck}] equations.
 {Correlations enter the BSK equation through a fourth-order gradient of the potential }[\emph{viz}.~\cref{Eq:mP}], {
hence, mainly affect regions where the potential varies the strongest.
Therefore, within BSK theory,  beyond-mean-field correlations presumably affect a system's capacitance }[\cref{eq:BSKcapacitance}]  {more than its resistance. }
If so, we can approximate the resistance by $R=2\lambda_D^2 L/(\varepsilon D)$, and the canonical setup should charge in response to a small applied potential step at a timescale \cite{zhao2011diffuse}
\begin{equation}\label{eq:tau2}
   \tau_{\mathrm{late}}=\frac{RC}{2}=\dfrac{\lambda_{D}L}{D}\dfrac{\sqrt{1+2\ell_c/\lambda_D}}{1+\ell_c/\lambda_D},
\end{equation}
which, for $\ell_c\ll\lambda_D$, reduces to Macdonald's timescale $\lambda_DL/D$.
In the opposite regime $\ell_c\gg\lambda_D$ of strong correlations, the relaxation time scales as $\tau_{\mathrm{late}}\propto L\lambda_D^{3/2}/(D\ell_c^{1/2})$.
Zhao also found this scaling, both numerically and through matched asymptotic approximations of the BSK-NP equations for $\lambda_D/L\to0$ \cite{zhao2011diffuse}.
Below, we show that \cref{eq:tau2} does not hold generally, but only when $\ell_c\ll L^{2/3}\lambda_D^{1/3}$.
The regime where $\ell_c\approx L^{2/3}\lambda_D^{1/3}$ corresponds to strongly correlated, strongly confined electrolytes, as may be realized in surface force balance experiments with ionic liquids, for example.

Other studies on the effect of correlations on EDL formation were mostly numerical and focused on applied potentials beyond the thermal voltage, the regime relevant to applications \cite{alijo2015effects, lee_prl_2015, yochelis_pccp_2020, gupta_prl_2020}.
Lee and coworkers found that, for large applied potentials, late-time charging goes with the timescale $L^2(\lambda_D/\ell_c)^{3/2}/D$ \cite{lee_prl_2015}, much slower than the $RC$ time \cref{eq:tau2}.
Notwithstanding such late-time charging slow down, the $RC$ relaxation mode could still be relevant to correlated electrolytes subject to large potentials, as EDLs acquire much of their charge at times comparable to $\lambda_D L/D$ \cite{ma_jcp_2022}.
Hence, a comprehensive analysis of the response of the canonical setup subject to a small applied potential---currently missing---is highly desirable. 
Here, we analyze the canonical setup for small applied potentials over the whole $\ell_c/\lambda_D$ range.
The system's charging time depends nonmonotonically on $\ell_c/\lambda_D$; in agreement with Zhao's results in the relevant parameter regime, but more complicated outside it.

\section{Model}
\subsection{Setup}
We consider a strong binary electrolyte, where $z_+$ and $z_-$ are the valencies and $X$ and $Y$ the stochiometric coefficients of the cations and anions.
The electrolyte has a salt concentration $c_0$ and temperature $T$, and is between two parallel flat blocking electrodes separated by a distance $2L$.
The Cartesian coordinate $x$ runs from the left ($x=-L$) to the right ($x=L$) electrode. 

\subsection{Governing equations}\label{sec:Model}
We study the response of the above cell to a small suddenly applied potential, in particular, its spatiotemporal ion concentrations $c_\pm(x,y,z,t)$ and electrostatic potential $\psi_\pm(x,y,z,t)$.
We model the relation between these variables through the BSK equation \cite{bazant_pre_2011,storey_pre_2012},
\begin{equation}\label{Eq:mP}
    \varepsilon(\ell_c^2\nabla^2-1)\nabla^2\psi =eq,
\end{equation}
where $q=z_+c_+ +z_- c_-$ is the local charge density, and $e$ is the proton charge.

The ionic densities satisfy the continuity equation 
\begin{equation}\label{eq:continuity}
    \partial_t c_{\pm} = -\nabla\cdot\vec{J}_{\pm},
\end{equation}
where we model the ionic fluxes $\vec{J}_{\pm}$ with the NP equation,
\begin{equation}\label{eq:nernstplanck}
    \vec{J}_{\pm}=-D\nabla c_{\pm}-Dc_{\pm}\dfrac{z_{\pm}e}{kT}\nabla \psi.
\end{equation} 
where $k$ is Boltzmann's constant and $D$ is the ionic diffusivity which, for simplicity, we consider to be the same for cations and anions.

When the electrodes are large compared to their separation, edge effects can be ignored, and $\psi, c_\pm$, and $J_\pm$ depend on the single spatial coordinate $x$ (and time $t$).
The governing equations \cref{Eq:mP,eq:continuity,eq:nernstplanck} then simplify to
\begin{subequations}\label{eq:base}
    \begin{align}\label{Eq:1DMP}
        \varepsilon\left( \ell_c^2\partial_x^2-1\right)\partial_x^2\psi&=eq,\\
        \partial_t c_{\pm} &= -\partial_x J_{\pm},\\
        J_{\pm}&=-D\partial_x c_{\pm}-Dc_{\pm}\dfrac{z_{\pm}e}{kT}\partial_x \psi.
    \end{align} 
\end{subequations}

\subsection{Initial and boundary conditions}
The electrolyte is uniform initially, 
\begin{subequations}\label{eq:initialcondition}
\begin{align}
    c_+ \big|_{x,t=0}&=Xc_0,\\
    c_- \big|_{x,t=0}&=Yc_0.
\end{align}    
\end{subequations}
At time $t=0$, we apply a potential difference $2\Psi$ between the blocking electrodes, so that
\begin{subequations}\label{eq:boundaryconditions}
    \begin{align}
        \psi\big|_{x=\pm L,t} &= \pm \Psi,\label{eq:bcpotential}\\
       J_\pm\big|_{x=\pm L,t} &= 0,\label{eq:bcnoflux}\\
       \ell_c\partial^3_x\psi\big|_{x=\pm L,t}&=0.\label{eq:bc3rdorder}
    \end{align}
\end{subequations}
Unlike \cref{eq:bcpotential,eq:bcnoflux}, the condition \eqref{eq:bc3rdorder} is not based on physical arguments; rather, with this choice, Ref.~\cite{storey_pre_2012} could capture correlation-induced effects on capacitance and streaming current data \footnote{We added the length  $\ell_c$ in \cref{eq:bc3rdorder} to ensure that, in the limit of uncorrelated electrolytes $\ell_c\to0$, our system of \cref{eq:base,eq:initialcondition,eq:boundaryconditions} reduces to those of Refs.~\cite{bazant_pre_2004,janssen_pre_2018}.}.
{Accordingly, \cref{eq:bc3rdorder} is the most widely-used additional boundary condition for the BSK equation.
De Souza and Bazant} \cite{desouza_jpcc_2020} derived a different boundary condition, $\ell_c\partial_x^3\psi\big|_{x=\pm L,t}=\pm\partial_x^2\psi\big|_{x=\pm L,t}$, based on mechanical equilibrium
{of an ideal solution [with zero excess chemical potential $\mu^{\mathrm{ex}}/(kT)=0$]. Concentration profiles from BSK theory with this new boundary condition agreed well with those from Grand Canonical Monte Carlo (GCMC) simulations \cite{valisko_aip_2018}, supporting the mechanical equilibrium route. However, the assumption $\mu^{\mathrm{ex}}/(kT)=0$ is not satisfied by GCMC simulations, especially in more correlated 2:1 or 3:1 electrolytes, where $\mu^{\mathrm{ex}}$ is comparable to the ideal chemical potential, see the Supplementary Information of Ref.~\cite{valisko_aip_2018}. 
Because of this concern, we use the common boundary condition \eqref{eq:bc3rdorder} in the main text, and de Souza and Bazant's \cite{desouza_jpcc_2020} boundary condition in \cref{app:A}.
} 

\Cref{eq:base,eq:initialcondition,eq:boundaryconditions} form a closed set that we can solve for $\psi(x,t)$ and $c_\pm(x,t)$.
From $\psi(x,t)$ we determine the surface charge density $\sigma$ using Gauss's law and \cref{eq:bc3rdorder}, 
\begin{equation}\label{eq:sigmaP}
    \sigma\big|_{x=\pm L} = \pm \varepsilon\partial_x \psi\big|_{x=\pm L},
\end{equation}
which, in turn, gives access to the transient electronic current per electrode area, $\iota(t)= \diff\sigma/\diff t$.

\subsection{Expansion for small applied potential}
We rewrite \cref{eq:base} in terms of the charge density $q$ and salt density $c_{\rm salt}=c_++c_-$ and corresponding fluxes $J_q=z_+J_++z_-J_-$ and $J_{\rm salt}=J_+ + J_-$.
For $q$, we find
\begin{subequations}\label{eq:base2}
    \begin{align}
    \partial_t q &= -\partial_x J_q,\\
    J_q&=-D\partial_x q-D(c_+z_+^2+c_-z_-^2)\dfrac{e}{kT}\partial_x \psi.\label{eq:contqq}
    \end{align} 
\end{subequations}
Likewise, for $c_{\rm salt}$ we find
\begin{equation}
    \partial_t c_{\rm salt}=D\partial^2_x c_{\rm salt}+D\partial_x\left(q\dfrac{e}{kT}\partial_x \psi\right).\label{eq:contsalt}
\end{equation} 
We consider a case of small applied potentials, $\Psi\ll kT/e$. 
We can then write asymptotic approximations to $q=q_0+\Phi q_1+\mathcal{O}(\Phi^2)$, $c_{\rm salt}=c_{\rm salt,0}+\Phi c_{\rm salt,1}+\mathcal{O}(\Phi^2)$, and $\psi=\psi_0+\Phi \psi_1+\mathcal{O}(\Phi^2)$ with $\Phi=e\Psi/k T$ a small parameter.
When no potential difference is applied, the electrolyte is not perturbed, so $q_0=0$, $\psi_0=0$, and $c_{\rm salt,0}=(X  + Y)c_0$.
Inserting the above expansions into \cref{eq:base2,eq:contsalt} yields 
\begin{subequations}
\begin{align}\label{eq:cont} \partial_{t}q_1&=D\partial^2_{x}q_1+D\dfrac{e}{kT}S\partial^2_{x}\psi_1+\mathcal{O}(\Phi^2),\\
    \partial_t c_{\rm salt,1} &=D\partial^2_x c_{\rm salt,1}+\mathcal{O}(\Phi^2),\label{eq:contsalt2}
\end{align}
\end{subequations}
where $S=Xc_0 z_+^2 + Yc_0 z_-^2$ is twice the ionic strength of the bulk electrolyte. \Cref{eq:contsalt2} shows that the salt concentration perturbation $c_{\rm salt,1}$ is governed by an ordinary diffusion equation, whose initial condition [$c_{\rm salt,1}(x,t)=0$] already satisfies its corresponding boundary conditions [$J_{\rm salt,1}(\pm L,t)=0$].
Hence, at $\mathcal{O}(\Phi)$ the salt density perturbation is trivial, $c_{\rm salt,1}(x,t)=0$. 
From hereon, we omit $\mathcal{O}(\Phi^2)$ terms and focus on $q_1$ and $\psi_1$, whose subscripts 1 we drop for readability.

\subsection{Dimensionless formulation}
We write $\tilde{t}=tD/L^2$, $\tilde{x}=x/L$, $\phi=e\psi/kT$, $\tilde{c}=c/S$, and $\tilde{q}=q/S$, and use three length scale ratios, $\gamma=\ell_c/L$, $\delta =\ell_c/\lambda_{D}$, and $\epsilon=\lambda_{D}/L$, where $\lambda_D=\sqrt{\varepsilon kT/(e^2 S)}$ is the Debye length.
Only two length scale ratios are independent---they satisfy $\gamma=\delta\epsilon$---but we use all three below to shorten expressions.
In terms of the dimensionless parameters, the governing \cref{Eq:1DMP,eq:cont} read
\begin{subequations}\label{Eq:start2eq}
\begin{align}
    \gamma^2\partial^4_{\tilde{x}}\phi -\partial^2_{\tilde{x}}\phi&=\epsilon^{-2}\tilde{q},\label{eq:start}\\
    \partial_{\tilde{t}}\tilde{q}&=\partial^2_{\tilde{x}}\tilde{q}+\partial^2_{\tilde{x}}\phi\label{eq:contt},
\end{align}
\end{subequations}
the initial condition \eqref{eq:initialcondition} reads
\begin{equation}\label{eq:initialcondition_dimless}
    \tilde{q}_\pm \big|_{\tilde{x},\tilde{t}=0}=0,
\end{equation}
and the boundary conditions \eqref{eq:boundaryconditions} read
\begin{subequations}\label{eq:bceq_dimless}
    \begin{align}
        \phi\big|_{\tilde{x}=\pm 1}&=\pm\Phi,\\
        \partial_{\tilde{x}} \tilde{q}+\partial_{\tilde{x}}\phi\big|_{\tilde{x}=\pm 1}&=0\label{eq:nofluxbc},\\
        \partial^3_{\tilde{x}}\phi\big|_{\tilde{x}=\pm 1}&=0.
    \end{align}
\end{subequations}
Combining \cref{eq:start,eq:nofluxbc} gives
\begin{equation}\label{eq:bcfifthorder}
    \gamma^2\epsilon^2\partial_{\tilde{x}}^5\phi+\partial_{\tilde{x}}\phi\big|_{\tilde{x}=\pm 1}=0.
\end{equation}
We scale the surface charge density and current by $\tilde{\sigma} =\sigma/(eL S)$ and $\tilde{\iota} =\iota L/(eD S)$, giving
\begin{equation}\label{eq:sigma_dimless}
    \tilde{\sigma}\big|_{\tilde{x}=\pm 1} = \pm \epsilon^{2}\partial_{\tilde{x}} \phi\big|_{\tilde{x}=\pm 1}     
\end{equation}
and $\tilde{\iota}=\diff\tilde{\sigma}/\diff \tilde{t}$.
We omit tildes from hereon for readability.

We estimate what $\gamma, \delta$, and $\epsilon$ are physically realistic by considering extreme values of $L, \lambda_D$, and $\ell_c$.
Electrode separations $L$ could range from $\sim\SI{e-2}{\meter}$ \cite{zhao_jpcc_2024} to $\SI{e-9}{\meter}$ (in the surface force balance apparatus, for example).
$\lambda_{D}$ ranges from $\sim\SI{e-10}{\meter}$ to $\SI{e-6}{\meter}$ for dense to dilute electrolytes.
Last, the correlation length $\ell_c$ is supposed to model correlations over several molecular diameters at most, implying that $\ell_c$ would range from $\sim\SI{e-10}{\meter}$ to $\sim\SI{e-8}{\meter}$ from atomic to organic ions.
We find $\gamma=[10^{-8},10], \delta=[10^{-4},10^{2}]$, and $\epsilon=[10^{-8},10^{3}]$ and note that central values are more realistic than values near the edges of these interval. 
To comprehensively study the ramifications of our BSK-NP model, we will also consider even larger $\delta$ below, as they may become accessible in the future. 

\section{Equilibrium}\label{sec:Equilibrium}
At equilibrium, \cref{eq:contt} reduces to $0=\partial^2_{x}q+\partial^2_{x}\phi$.
To eliminate $q$, we insert \cref{eq:start}, yielding 
\begin{equation}\label{eq:4thode}
    \gamma^2\partial_x^6 \phi-\partial_x^4 \phi+\epsilon^{-2}\partial_x^2 \phi =0.
\end{equation}
We rewrite the characteristic polynomial equation $\gamma^2r^6-r^4+\epsilon^{-2}r^2=0$ of \cref{eq:4thode} with $\vartheta=r^2$ to
\begin{equation}\label{eq:char}
    \vartheta(\gamma^2\vartheta^2-\vartheta+\epsilon^{-2})=0.
\end{equation}
\Cref{eq:char} is solved by $\vartheta=0$ and $\vartheta_{\pm}=\left(1\pm\sqrt{1-4\delta^2}\right)/(2\gamma^2)$.
The sign of the discriminant, $1-4\delta^2$, changes at $\delta=1/2$, so the solution to \cref{eq:4thode} takes different forms depending on $\delta$.

For ${\delta<1/2}$, \cref{eq:char} has real roots at $\vartheta_{\pm}=\left(1\pm\sqrt{1-4\delta^2}\right)/(2\gamma^2)$.
\Cref{eq:4thode} is solved by $\phi(x)=A_1+A_2x+A_3\exp(r_{-}x)+A_4\exp(-r_{-}x) +A_5\exp(r_{+}x)+A_6\exp(-r_{+}x)$, with
$r_{\pm}=\left[\left(1\pm\sqrt{1-4\delta^2}\right)/(2\gamma^2)\right]^{1/2}$.
We fix $A_1,\ldots, A_6$ by \cref{eq:bceq_dimless,eq:bcfifthorder}, yielding
\begin{align}\label{eq:solsm}
   \frac{\phi(x)}{\Phi}&=\dfrac{\sinh(r_{-}x)}{\sinh r_{-}-\cosh r_{-} \tanh(r_{+})r_{-}^3/r_{+}^3}\nn
   &\quad+\dfrac{\sinh(r_{+}x)}{\sinh r_{+}-\tanh r_{-}\cosh(r_{+})r_{+}^3/r_{-}^3}.
\end{align}
For $\gamma\rightarrow 0$, the roots tend to $r_{-}\to 1/\epsilon$ and $r_{+}\to\infty$, and \cref{eq:solsm} reduces to the known expression $ \phi(x)=\Phi\sinh(x/\epsilon)/\sinh (1/\epsilon)$ \cite{janssen_pre_2018}.

For $\delta=1/2$, \cref{eq:char} has real roots at $\vartheta=1/(2\gamma^2)$, giving double roots at $r=1/(\sqrt{2}\gamma)$ and at $-r$. 
\Cref{eq:4thode} is solved by $\phi(x)=A_1+A_2x+A_3\exp(r x)+ A_4x\exp(r x)+A_5\exp(-r x)+A_6x\exp(-r x)$.
Fixing $A_1,\ldots, A_6$ by \cref{eq:bceq_dimless,eq:bcfifthorder} yields 
\begin{equation}\label{eq:soleq}
    \frac{\phi(x)}{\Phi}=\dfrac{\sinh(r x)[r\sinh r+3\cosh r]-r x\cosh(r x)\cosh r}{3\sinh r\cosh r-r}.
\end{equation}

\begin{figure}
    \centering
	\includegraphics[width = \linewidth]{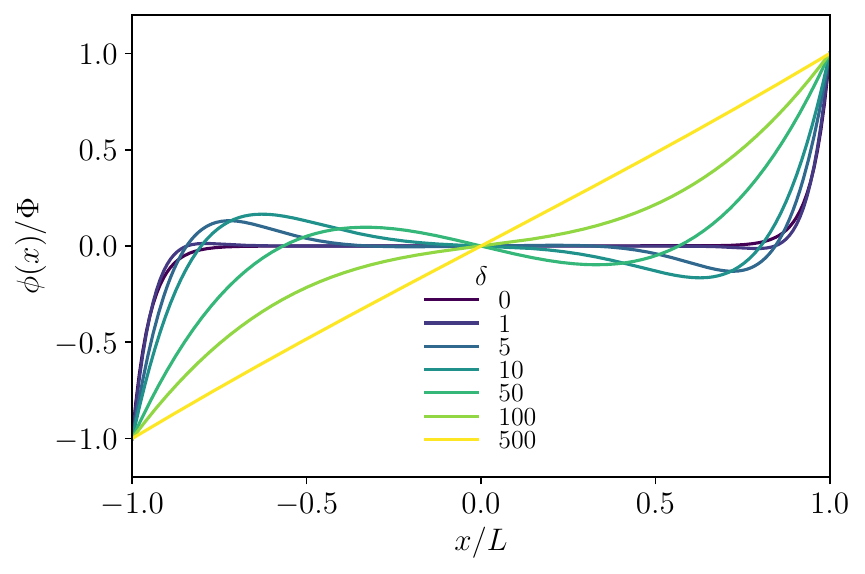}
    \caption{Equilibrium potential $\phi(x)/\Phi$ as predicted by \cref{eq:solsm,eq:soleq,eq:sollmcomplete} vs. the distance $x$ of the electrode for $\epsilon=1/20$ and several $\delta$.}\label{fig:equilibrium}
\end{figure}

For $\delta>1/2$, \cref{eq:char} has complex roots at  $\vartheta_{\pm}=\left(1\pm i\sqrt{4\delta^2-1}\right)/(2\gamma^2)$, giving $r_1=v+iw$, $r_2=-r_1$, $r_3=v-iw$, and $r_4=-r_3$, where $v=\sqrt{2\delta+1}/(2\gamma)$ and $w=\sqrt{2\delta-1}/(2\gamma)$.
\Cref{eq:4thode} is solved by $\phi(x)=A_1+A_2x+\exp(vx)[A_3\cos(wx)+A_4\sin(wx)]+ \exp(-vx)[A_5\cos(wx)+A_6\sin(wx)]$.
Fixing the constants, we find
\begin{subequations}\label{eq:sollmcomplete}
\begin{align}\label{eq:sollm}
    &\frac{\phi(x)}{\Phi}=\dfrac{2}{\Xi}\Big\{\big[w(w^2-3v^2)\cosh v\cos w\nn
    &\qquad\qquad-v(v^2-3w^2)\sinh v\sin w\big]\sinh(vx)\cos(wx)\nn
    &\qquad\qquad+\big[v(v^2-3w^2)\cosh v \cos w\nn
    &\qquad\qquad+w(w^2-3v^2)\sinh v\sin w\big]\cosh(vx)\sin(wx)\big\},
    \intertext{where}
    &\Xi=w(w^2-3v^2)\sinh (2v)+v(v^2-3w^2)\sin (2w).
\end{align}
\end{subequations}
\Cref{fig:equilibrium} shows the potential $\phi(x)$ for various $\delta$ as predicted by \cref{eq:solsm,eq:soleq,eq:sollmcomplete}.
This figure is similar to Fig.~1(a) in Ref.~\cite{storey_pre_2012}.
That article considered the EDL near a single electrode, fully characterized by $\delta$, while our curves concern two electrodes, which depend on two length scale ratios, $\epsilon$ and $\delta$.
As in Ref.~\cite{storey_pre_2012}, we observe overscreening for all considered $\delta$ values except $\delta=0$.

\section{Transient response}\label{sec:nonEquilibrium}
\subsection{Solution in Laplace domain}
We analyze the cell's transient response by Laplace-transforming the governing \cref{Eq:start2eq},
\begin{subequations}
    \begin{align}
    \gamma^2\partial^4_x\hat{\phi} -\partial^2_x\hat{\phi}&=\epsilon^{-2}\hat{q},\label{eq:laplbsk}\\
    s\hat{q}-q(x,t=0) &= \partial_x^2 \hat{q}+\partial_x^2 \hat{\phi},\label{eq:laplcont}
    \end{align}
\end{subequations}
where we denote the time-domain Laplace transform of a function $f(x,t)$ by $\hat{f}(x,s)=\int\limits_0^\infty f(x,t)\exp(-st) \diff t$.
In \cref{eq:laplcont}, $q(x,t=0)$ drops because of \cref{eq:initialcondition_dimless}.
The boundary conditions \cref{eq:bceq_dimless,eq:bcfifthorder} turn into
\begin{subequations}\label{eq:bceq_dimless_laplace}
    \begin{align}
        \hat{\phi}\big|_{\tilde{x}=\pm 1}&=\pm\frac{\Phi}{s},\\
        \gamma^2\epsilon^{2}\partial_{\tilde{x}}^5\hat{\phi}+\partial_{\tilde{x}}\hat{\phi}\big|_{\tilde{x}=\pm 1}&=0,\\
        \partial^3_x\hat{\phi}\big|_{\tilde{x}=\pm 1}&=0.
    \end{align}
\end{subequations}
We eliminate $\hat{q}$ from \cref{eq:laplcont} with \cref{eq:laplbsk}, 
\begin{equation}\label{Eq:4Poiss}
    \gamma^2\partial_x^6\hat{\phi}-(\gamma^2s+1)\partial_x^4\hat{\phi}+(\epsilon^{-2}+s)\partial_x^2\hat{\phi}=0,
\end{equation}
whose solution reads $\hat{\phi}(x,s) = A_1+A_2x+A_3\exp(gx)+A_4\exp(-gx)+A_5\exp(hx)+A_6\exp(-hx)$, where
\begin{subequations}\label{eq:gandh}
\begin{align}
    &g=\sqrt{\dfrac{\gamma^2s+1-\sqrt{\left(\gamma^2s-1\right)^2-4\delta^2}}{2\gamma^2}},\label{eq:g}\\
    &h=\sqrt{\dfrac{\gamma^2s+1+\sqrt{\left(\gamma^2s-1\right)^2-4\delta^2}}{2\gamma^2}}.
\end{align}
\end{subequations}
Fixing $A_1,\ldots, A_6$ by \cref{eq:bceq_dimless_laplace} gives
\begin{subequations}\label{Eq:laplphifull}
\begin{align}
    \label{Eq:laplphi}
    &\hat{\phi}(x,s)=\dfrac{\Phi}{\Omega s}\bigg\{\dfrac{h^3\sinh(gx)}{\cosh g}-\dfrac{g^3\sinh(hx)}{\cosh h}\nn
    &\qquad\qquad\qquad-gh(g^2-h^2)\left[(gh\gamma\epsilon)^{2}-1\right]x\bigg\},
    \intertext{where}
    &\Omega= h^3\tanh g-g^3\tanh h-gh(g^2-h^2)\left[(gh\gamma\epsilon)^{2}-1\right].\label{eq:Omega}
\end{align}    
\end{subequations}
For $\gamma\to0$, \cref{Eq:laplphifull} reduces to the dimensionless form of Eq.~(12) of Ref.~\cite{janssen_pre_2018}.

The transient response of the cell is governed by~\cite{janssen_pre_2018}
\begin{equation}\label{eq:Res}
    \mathcal{L}^{-1}\left\{\hat{\phi}(x,s)\right\}= \sum_{s^\star}\text{Res}\left(\hat{\phi}(x,s)e^{ts},{s^\star}\right),
\end{equation}
where $s^\star$ are the poles of $\hat{\phi}(x,s)$.
Except for the pole at $s=0$, the poles of $\hat{\phi}(x,s)$ coincide with the zeros of $\Omega$.
\Cref{eq:Res} shows that the location of these poles determine the relaxation times $\tau^\star=-1/s^\star$ of the different modes. 
$s=0$ sets the steady-state potential $\phi(x,t\to\infty)$.
Likewise, the late-time response of the cell towards the equilibrium state is set by the pole $s_{\rm late}$ closest to $s=0$ on the negative $s$ axis, decaying exponentially with a timescale $\tau_{\rm late}=-1/s_{\rm late}$.

We have not found manageable analytical expressions for the residues in \cref{eq:Res}, but the poles $s^\star$ do yield to analytical study.
We discuss numerical results for $\phi(x,t)$, $\iota(t)$ and their late-time relaxation time $\tau_{\rm late}$ in \cref{sec:numerical}.
In \cref{sec:poles} we study the poles $s^\star$ of $\hat{\phi}(x,s)$ and find analytical approximations to $s_{\rm late}$ and  $\tau_{\rm late}$.

\begin{figure}
    \centering
	\includegraphics[width = \linewidth]{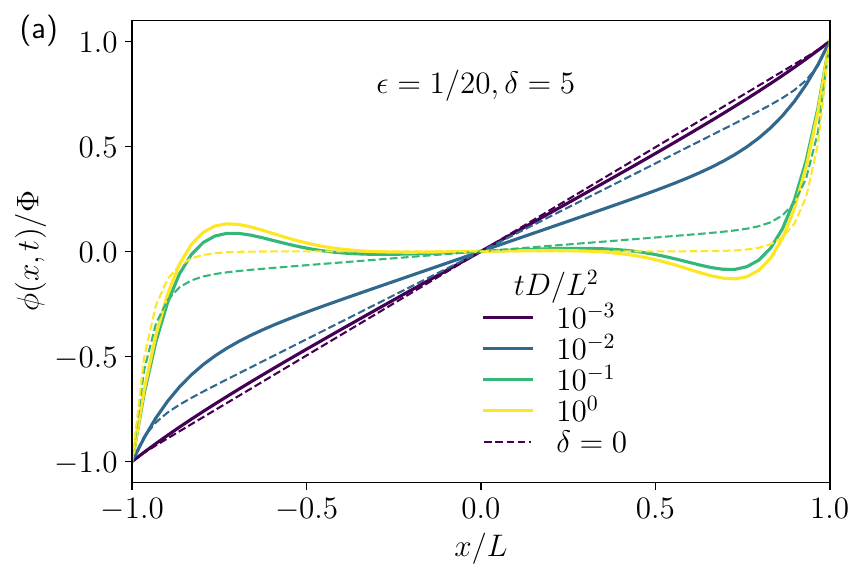}
    \includegraphics[width = \linewidth]{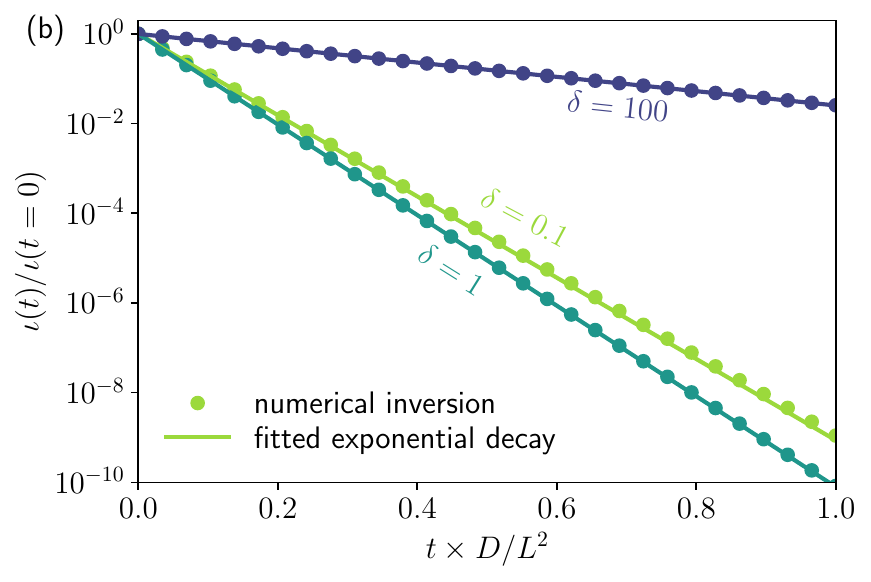}
    \caption{(a) Transient potential $\phi(x,t)/\Phi$ for $\epsilon=1/20$ and $\delta=5$ (lines) and $\delta=0$ (dashed lines) at several times after applying a step potential at $t=0$, as found by numerically Laplace inverting \cref{Eq:laplphifull}. 
    (b) Transient electric current $\iota$ scaled to its initial value (dots), at $\epsilon=1/20$ and  $\delta=0.1, 1$, and $100$, determined by numerical evaluation of \cref{eq:electriccurrent}. Lines indicate fitted exponential decays to these data.}\label{fig:transient_phi_and_I}
\end{figure}

\subsection{Numerical results for $\phi(x,t), \iota(t)$, and $\tau_{\rm late}$}\label{sec:numerical}
\Cref{fig:transient_phi_and_I}(a) shows $\phi(x,t)$ for $\epsilon=1/20$ and $\delta=5$ (lines) and $\delta=0$ (dashed lines), as determined by numerically evaluating  $\mathcal{L}^{-1}\big\{\hat{\phi}(x,s)\big\}$, with $\hat{\phi}(x,s)$ from \cref{Eq:laplphifull}.
We see that the $\phi(x,t)$ of these two $\delta$ differ throughout the transient relaxation.

Next, the areal electronic current amounts to
\begin{equation}\label{eq:electriccurrent}
    \iota(t)= \mathcal{L}^{-1}\big\{\hat{\iota}(s)\big\}= \mathcal{L}^{-1}\big\{s\hat{\sigma}\big\},
\end{equation}
where we used $\hat{\iota}(s)=\mathcal{L}\left\{\diff\sigma/\diff t\right\}=s\hat{\sigma}-\sigma(t=0)$ and dropped $\sigma(t=0)$. 
We numerically determined \cref{eq:electriccurrent} using 
\begin{align}\label{eq:sigmalapl}
    \hat{\sigma}(s)=\dfrac{\Phi}{\Omega s} \gamma^2\epsilon^{4}(g^3 h^5-g^5 h^3),
\end{align} which follows from inserting \cref{Eq:laplphifull} into  Gauss's law [\cref{eq:sigma_dimless}].
\Cref{fig:transient_phi_and_I}(b) shows $\iota(t)$ scaled to its initial value, $\iota(t=0)$, vs. time for $\epsilon=1/20$ and  $\delta=0.1, 1$, and $100$.
We see that $\iota(t)$ decays exponentially with a timescale $\tau_{\rm late}$ that depends nonmonotonically on $\delta$: $\tau_{\rm late}$ first decreases and then increases with increasing $\delta$.
We determine $\tau_{\rm late}$ by fitting exponentially decaying functions to the data in \cref{fig:transient_phi_and_I}(b), indicated there with lines.

\begin{figure}
	\centering
	\includegraphics[width = 0.98\linewidth]{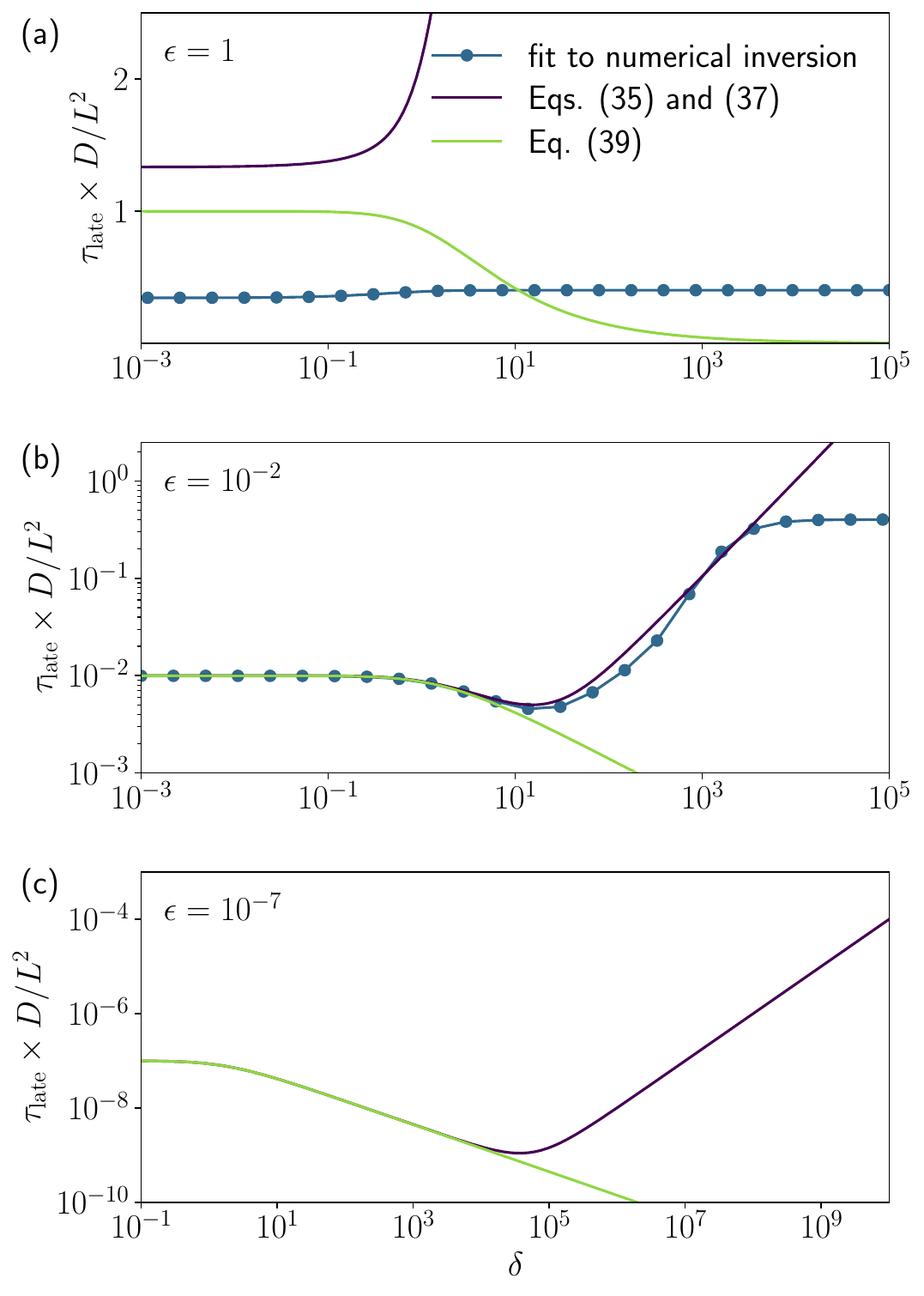}
    \caption{Relaxation time as determined by fitting the exponential decay of \cref{fig:transient_phi_and_I}. Also shown are $\tau_{\mathrm{late}}$ predictions from \cref{eq:tau_solution,eq:sol} (purple) and \cref{eq:tau_late} (green) for (a) $\epsilon=1$, (b) $\epsilon=10^{-2}$, and (c) $\epsilon=10^{-7}$.}\label{fig:tau}
\end{figure}

We repeated those fits for many $\delta$;
\Cref{fig:tau}(a) and (b) shows results for $\tau_{\rm late} D/L^2$  vs. $\delta$ (blue) for $\epsilon=1$ and $1/100$.
The log-log plot \cref{fig:tau}(b) shows that $\tau_{\rm late}$ depends nonmonotonically on $\delta$.
$\tau_{\rm late}$ is minimal around $\delta\approx20$, where it is about four times smaller than for $\delta=0$.
For $\delta>\epsilon^{-2}$, $\tau_{\rm late}$ approaches a plateau whose height lies around $4/\pi^2=0.4053$.
We studied more $\epsilon$ values and saw that the height of the plateau does not depend on $\delta$ and $\epsilon$ (not shown).
Note that the $\tau_{\rm late}$ minimum and part of the subsequent increase occur for $\delta\sim10^2$ and $\epsilon=10^{-2}$ that should be experimentally accessible.
Conversely, the plateau occurs at $\delta$ that are physically unrealistic now, but may be reached in the future.

\subsection{Analytical approximations for $\tau_{\rm late}$}\label{sec:poles}
To understand the data in \cref{fig:tau} better, we seek analytical expressions for $\tau_{\rm late}$.
The relaxation times of $\iota(t)$ are set, through \cref{eq:electriccurrent}, by the poles $s^\star$ of $\hat{\sigma}(s)$.
Their locations coincide with the zeros of $\Omega$, that is, the solutions of
\begin{equation}\label{eq:pole}
    h^3\tanh g-g^3\tanh h=gh(g^2-h^2)\left[(gh\gamma\epsilon)^{2}-1\right].
\end{equation}
$\Omega$ and its zeros depend intricately on $\gamma, \delta$, and $\epsilon$.
In our analytical derivation below, we will focus on the nonoverlapping EDL regime, $\epsilon\ll1$, as it is more practically relevant than $\epsilon\sim1$ and smaller. 
We keep $\epsilon$ fixed, and study the effect of electrostatic correlations by varying $\delta$ (which is equivalent to varying $\gamma$).

\begin{figure}
	\centering
    \includegraphics[width = \linewidth]{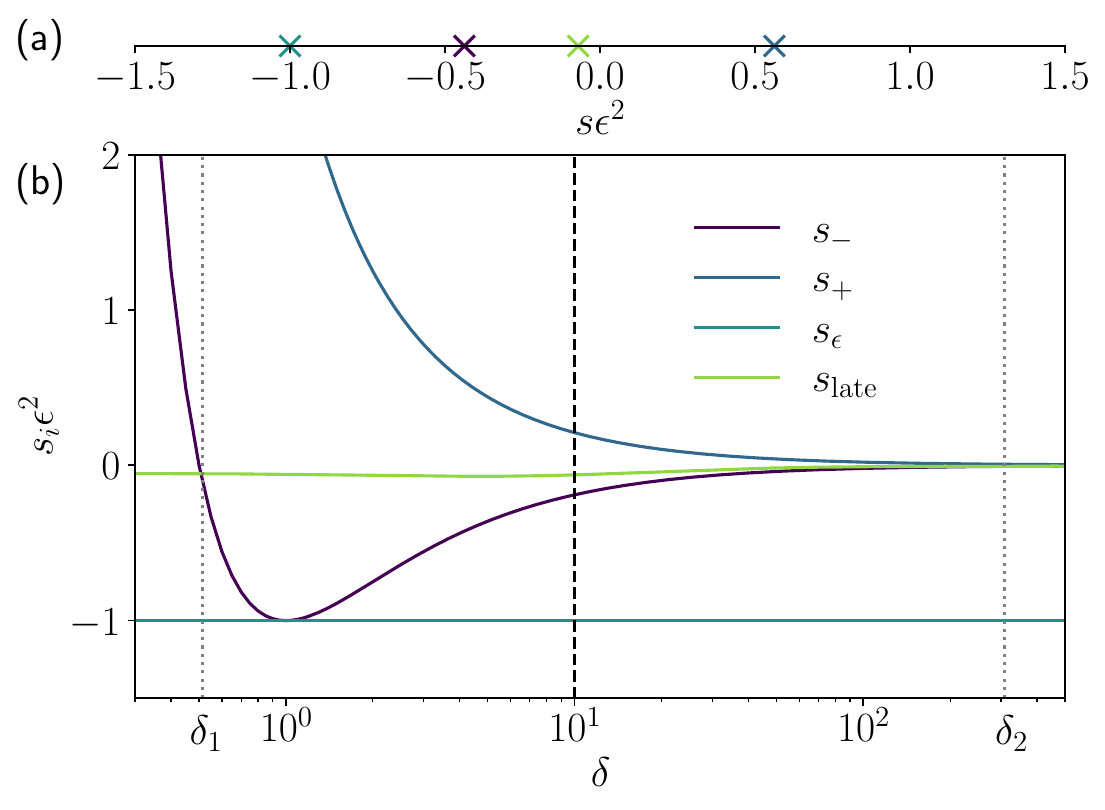}    
    \caption{(a) The location of the poles for $\delta=10$ and $\epsilon=1/20$ on the $s$-axis scaled with $\epsilon^2$.
    (b) The location of the poles $s_i$ scaled with $\epsilon^2$ as a function of $\delta$ for $\epsilon=1/20$. The dotted grey lines show $\delta_1$ and $\delta_2$. The dashed line shows $\delta=10$, corresponding to panel (a).}\label{fig:polestructure}
\end{figure}

We graphically analyzed $1/\Omega$ in the complex $s$ plane and did not observe any poles other than on the real $s$ axis.
Poles at complex $s$ would lead to oscillating potentials $\phi(x,t)$ and currents $\iota(t)$; no such oscillations are visible in the numerical Laplace inversions presented in \cref{fig:transient_phi_and_I}.
Hence, we assert that all poles of $1/\Omega$ lie on the real $s$ axis, and focus on this axis from hereon.

\Cref{fig:polestructure}(a) shows the location of several poles on the $s$ axis near $s=0$ for $\epsilon=1/20$ and $\delta=10$ \footnote{We determined $s_+,s_-$, and $s_{\epsilon}$ analytically and $s_{\mathrm{late}}$ numerically.};
\Cref{fig:polestructure}(b) shows how these locations shift when $\delta$ is varied
\footnote{For $\delta=0$, our model reduces to the one in Ref.~\cite{janssen_pre_2018}, and the poles are identical to those of that paper.}.
First, the poles $s_{\epsilon},s_+$, and $s_-$ correspond to trivial solutions to \cref{eq:pole} at $g=0$, $h=0$, and $g=h$. 
Specifically, both $g=0$ and $h=0$ give $s_{\epsilon}=-\epsilon^{-2}$ (dark cyan cross and line in \cref{fig:polestructure}), while $g=h$ yields two solutions, $s_{\pm}=1/(\delta\epsilon)^2\pm 2/(\epsilon^2\delta)$ (purple and dark blue crosses and lines in \cref{fig:polestructure}).
We calculated the residues of these trivial solutions numerically and found that they are zero; hence, they do not contribute to $\phi(x,t)$ and $\iota(t)$.
Second, $s_{\mathrm{late}}$ is the pole closest to $s=0$ with a nonzero residue (green curve in \cref{fig:polestructure}); we seek approximations to the location of this pole below, as it sets the late-time relaxation of our cell.
Third, there are infinitely many poles further on the negative $s$-axis. 
These poles give fast-decaying contributions to $\sigma(t)$, hence, only affect the early-time response of the cell.
 {We do not show these poles in }\cref{fig:polestructure}  {and do not consider them further, but we briefly discuss the system's early time response in }\cref{sec:early}  {using asymptotic approximations to} \cref{eq:sigmalapl}  {for $s\to-\infty$.}

Even though the pole $s_{-}$ does not contribute to $\phi(x,t)$ or $\iota(t)$, its location relative to $s_{\mathrm{late}}$ is important in our discussion of $s_{\mathrm{late}}$ below.
\Cref{fig:polestructure} shows that, for $\epsilon=1/20$, $s_{\mathrm{late}}$ and $s_{-}$ cross at $\delta_1\approx0.51$ and $\delta_2\approx308$.
As we do not know yet how $s_{\mathrm{late}}$ depends on $\delta$ and $\epsilon$, we cannot solve $s_{\mathrm{late}}=s_{-}$ yet for $\delta_1$ and $\delta_2$ (for a given $\epsilon$).

\begin{figure}
	\centering
	\includegraphics[width = \linewidth]{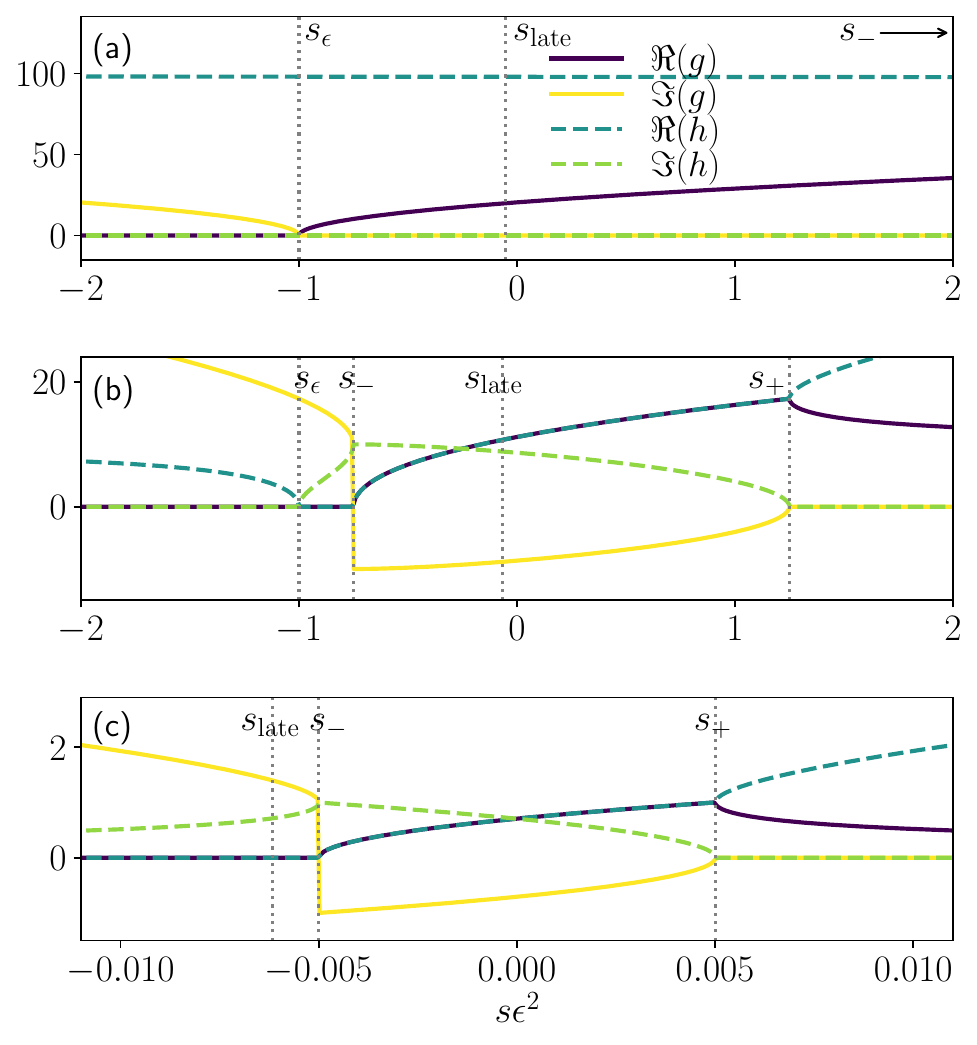}
    \caption{The real and imaginary parts of $g$ and $h$ for $\epsilon=1/20$, for (a) $\delta=0.2$ and $\gamma=0.01$, (b) $\delta=2$ and $\gamma=0.1$, and (c) $\delta=400$ and $\gamma=20$. The locations of $s_{\epsilon}$, $s_{\pm}$, and $s_{\mathrm{late}}$ are denoted with dotted lines.}\label{fig:realimag_gh}
\end{figure}

\subsubsection{Late-time response response for $\delta<\delta_1<\delta_2$}
For $\delta<\delta_1$, $s_{\mathrm{late}}$ lies on the interval $[s_{\epsilon},s_-]$.
\Cref{fig:realimag_gh}(a) shows that $g$ and $h$ are real on this interval.
For nonoverlapping EDLs (${\epsilon\ll1}$), we then find that $g\gg1$ and $h\gg 1$, thus $\tanh(g)\approx1$ and $\tanh(h)\approx1$, and \cref{eq:pole} simplifies to
\begin{equation}\label{eq:orig}
    h^3-g^3=gh(g^2-h^2)\left[(gh\gamma\epsilon)^{2}-1\right].
\end{equation}
We divide both sides by $h-g$, square both sides, use the identities $g^2+h^2=s+1/\gamma^2$,  $gh=\sqrt{\epsilon^{-2}+s}/\gamma$, and $g^2h^2\gamma^2-\epsilon^{-2}=s$, and introduce $m^2=\epsilon^{-2}+s$, to obtain
\begin{align}
    &(\gamma^2m^2-\delta^{2}+1+\gamma m)^2=\nn &\qquad \qquad m^2(1-m^2\epsilon^{2})^2(\gamma^2m^2-\delta^{2}+1+2\gamma m).
\end{align}
We divide by $\gamma^2m^2-\delta^{2}+1+2\gamma m$ and take square roots on both sides, and keep only the real solution \footnote{For $\gamma\rightarrow 0$, the left-hand side of \cref{eq:needtosolve} tends to 1, reducing to an equation at the bottom of page 4 of Ref.~\cite{janssen_pre_2018}.},
\begin{equation}\label{eq:needtosolve}
    \dfrac{\gamma^2m^2-\delta^{2}+1+\gamma m}{\sqrt{\gamma^2m^2-\delta^{2}+1+2\gamma m}}-m\left(1-m^2\epsilon^{2}\right)=0.
\end{equation}

\begin{figure}
	\centering
	\includegraphics[width = \linewidth]{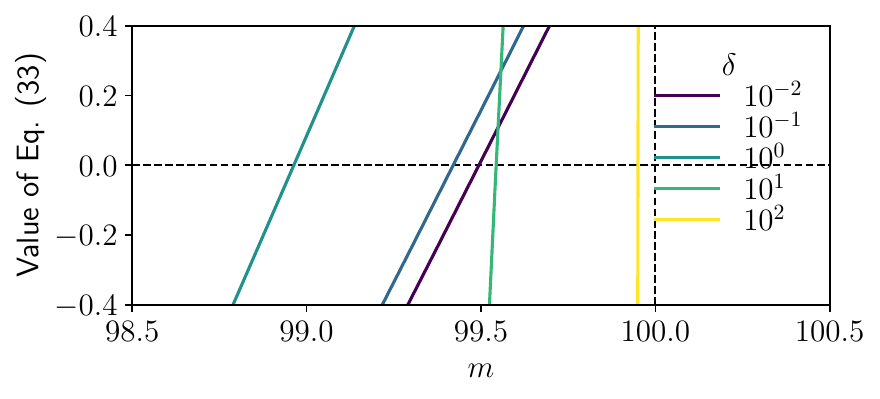}
    \caption{The left hand side of \cref{eq:needtosolve} for $\epsilon=10^{-2}$ and several $\delta$. 
    The zeros of these curves lie near $1/\epsilon$. The black dashed lines at $x=0$ and $y=1/\epsilon$ are guides to the eye.}\label{fig:sol_in_m}
\end{figure}
 
\Cref{fig:sol_in_m} shows the left-hand-side of \cref{eq:needtosolve} vs. $m$ for several $\delta$.
Hence, zeros in that graph correspond to solutions to \cref{eq:needtosolve}.
The dark blue curve, corresponding to $\delta=0.1<\delta_1$, is zero for $m\lessapprox \epsilon^{-1}$. 
Hence, we write $m=\epsilon^{-1}(z+1)$, where $z$ is small ($|z|\ll1$).
In terms of this yet-to-be-determined $z$, we find
\begin{equation}\label{eq:tau_solution}
    \tau_{\mathrm{late}}=-\frac{1}{s_{\mathrm{late}}}=-\frac{1}{m^2-\epsilon^{-2}}=-\frac{\epsilon^2}{z^2+2z}.
\end{equation}  
To find $z$, we insert $m=\epsilon^{-1}(z+1)$ into \cref{eq:needtosolve}, bring the square root term to the right-hand side, and square both sides, to find
\begin{align}\label{eq:zsq}
   & \{\delta^2[(z+1)^2-1]+1+\delta(z+1)\}^2=\nn
   &\qquad \epsilon^{-2}(z+1)^2[1-(z+1)^2]^2\nn
   &\qquad \times [\delta^2(z+1)^2-\delta^2+1+2\delta(z+1)].
\end{align}
Note that with the squaring, we introduce spurious solutions for $z$ in \cref{eq:zsq}.
By working out the brackets and neglecting $\mathcal{O}(z^3)$ terms, we find a quadratic equation for $z$, whose valid solution reads 
\begin{equation}\label{eq:sol}
    z=-\dfrac{\epsilon(1+\delta)}{2\sqrt{1+2\delta-\epsilon^{2}(\delta^2- 5\delta^3)/2}+\epsilon\delta(1+2\delta)}.
\end{equation}    
The combination of \cref{eq:tau_solution,eq:sol} sets the setup's late-time relaxation.
We show below that these two equations are in remarkable agreement with the numerical data presented in \cref{sec:numerical}, and as such, they form the key analytical result of this article.

In the argument leading up to \cref{eq:orig} we assumed $\epsilon\ll1$, so we expand the above expression,  
\begin{equation}
    z=-\epsilon\dfrac{1+\delta}{2\sqrt{1+2\delta}}+\epsilon^2\frac{\delta(1+\delta)}{4}+\mathcal{O}\left(\epsilon^3\right).
\end{equation}  
Inserting this into \cref{eq:tau_solution} gives 
\begin{equation}\label{eq:tau_late}
    \tau_{\mathrm{late}}=\epsilon \dfrac{\sqrt{1+2\delta}}{1+\delta}+\mathcal{O}\left(\epsilon^2\right)+\mathcal{O}\left(\epsilon^2\delta\right).
\end{equation}
Restoring units, the lowest order term reduces to \cref{eq:tau2}.

We plot $\tau_{\mathrm{late}}$ as predicted by \cref{eq:tau_solution,eq:sol} (purple) and \cref{eq:tau_late} (green), as well as the numerical data (blue dotted lines) discussed above, for $\epsilon=1$ [\cref{fig:tau}(a)] and $10^{-2}$ [\cref{fig:tau}(b)].
For overlapping EDLs ($\epsilon=1$), the analytical predictions do not capture the numerical data. 
This was to be expected, as we assumed $\epsilon\ll1$ in the argument leading up to \cref{eq:orig}.
For nonoverlapping EDLs ($\epsilon=10^{-2}$), both analytical expressions excellently capture $\tau_{\mathrm{late}}$ for $\delta$ even much beyond $\delta_1\approx0.51$, the regime considered here.
The next paragraph shows why that is.

\subsubsection{Late-time response response for $\delta_1<\delta<\delta_2$}
For $\delta_1<\delta<\delta_2$, $s_{\mathrm{late}}$ lies on the interval $[s_-,s_+]$.
\Cref{fig:realimag_gh}b shows that $g$ and $h$ are complex conjugates on that interval, $h=g^*$.
We write $g$ as 
\begin{align}\label{eq:ghcomplex}
    g&=\sqrt{\dfrac{\sqrt{\epsilon^{-2}+s}}{2\gamma}+\dfrac{\gamma^2s+1}{4\gamma^2}}+i\sqrt{\dfrac{\sqrt{\epsilon^{-2}+s}}{2\gamma}-\dfrac{\gamma^2s+1}{4\gamma^2}}
\end{align}
Introducing $g_r=\Re(g)$ and $g_i=\Im(g)$, we have $g=g_r+ig_i$ and $h=g_r-ig_i$. 
\Cref{eq:pole} is then
\begin{align}\label{eq:complexpole}
    &(g_r-ig_i)^3\tanh(g_r+ig_i)-(g_r+ig_i)^3\tanh(g_r-ig_i)=\nn
    &\qquad(g_r+ig_i)(g_r-ig_i)\left[(g_r+ig_i)^2(g_r-ig_i)^2\gamma^2\epsilon^{2}-1\right]\nn
    &\qquad\times[(g_r+ig_i)^2-(g_r-ig_i)^2]
\end{align}
Using $\tanh(a+ib)=[\sinh(2a)+i\sin(2b)]/[\cosh(2a)+\cos(2b)]$, we find
\begin{align}
    &i2\epsilon^{-2}\dfrac{(g_i^3-3g_r^2g_i)\sinh(2g_r)+g_i(g_r^2-2g_i^2)\sin(2g_i)}{\cosh(2g_r)+\cos(2g_i)}=\nn
    &\qquad i4g_rg_i(g_r^2+g_i^2)(g_r^2\gamma+g_i^2\gamma-\epsilon^{-1})(g_r^2\gamma+g_i^2\gamma+\epsilon^{-1})
\end{align}
As $\epsilon\ll1$, we have $\tanh(2g_r)=1$, so that
\begin{align}\label{eq:imag}    
    i2g_i(g_i^2-3g_r^2)&=i4g_rg_i(g_r^2+g_i^2)\left(g_r^2\gamma\epsilon+g_i^2\gamma\epsilon-1\right)\nn
    &\quad\times\left(g_r^2\gamma\epsilon+g_i^2\gamma\epsilon+1\right).
\end{align}
Inserting \cref{eq:ghcomplex} to \cref{eq:imag}, squaring (thereby introducing a spurious solution), and simplifying yields, in terms of $m=\sqrt{\epsilon^{-2}+s}$,
\begin{align}\label{eq:mmm}
    &\dfrac{\gamma^2(m^2-\epsilon^{-2})+1-2\gamma m}{[\gamma^2(m^2-\epsilon^{-2})+1]^2-4\gamma^2m^2}\left[\gamma^2(m^2-\epsilon^{-2})+1+\gamma m\right]^2\nn
    &\quad= m^2(m^2\epsilon^{2}-1)^2.
\end{align}
Simplifying the fraction, taking the square root of both sides of \cref{eq:mmm}, and using the negative value of the left hand side (the positive square root leads to a false solution), we find \cref{eq:needtosolve} as before.
Hence, up to $\delta_2$, which lies near $\epsilon^{-2}$, the late-time transient response is set by the solution to \cref{eq:needtosolve}.
This explains why we observed that \cref{eq:tau_solution,eq:sol} decently captured numerical data for $\tau_{\rm late}$ in \cref{fig:tau}(b) up to large values of $\delta\approx\epsilon^{-2}$.

Zhao \cite{zhao2011diffuse} focused on cases where $\delta\gg1$, for which \cref{eq:tau_late} reduces to 
\begin{align}\label{eq:zhaoscaling}
\tau_{\mathrm{late}}=\epsilon \sqrt{2/\delta}+\mathcal{O}\left(\epsilon\delta^{-3/2}\right)+\mathcal{O}\left(\epsilon^2\right)+\mathcal{O}\left(\epsilon^2\delta\right),
\end{align}
In dimensional units, the leading order term reads $\tau_{\mathrm{late}}=\sqrt{2}\lambda_{D}^{3/2}L/(D\ell_c^{1/2})$, in agreement with Ref.~\cite{zhao2011diffuse}.
\Cref{eq:zhaoscaling} predicts $\tau_{\mathrm{late}}$ to monotonically decrease with increasing $\delta$, contradicting our observations in \cref{sec:numerical} that $\tau_{\rm late}$ varies nonmonotonically with $\delta$.
When we instead consider \cref{eq:sol} for $\delta\gg1$, we find $z=-1/(2\delta)+\mathcal{O}(\delta^{-3/2})$. 
Inserting into \cref{eq:tau_solution} gives
\begin{equation}\label{eq:ourscaling}
\tau_{\mathrm{late}}=\epsilon^2\delta +\mathcal{O}(\epsilon^2),    
\end{equation}
which increases with $\delta$.
Restoring units, the leading order reads term $\tau_{\mathrm{late}}=\lambda_{D}\ell_c/D$.

We again study $\tau_{\mathrm{late}}$ as predicted by \cref{eq:tau_solution,eq:sol} (purple) and \cref{eq:tau_late} (green), now for $\epsilon=10^{-2}$ [\cref{fig:tau}(b)] and $10^{-7}$ [\cref{fig:tau}(c)], to see for what $\delta$ the scalings \cref{eq:zhaoscaling,eq:ourscaling} hold.
Unlike \cref{fig:tau}(b), \cref{fig:tau}(c) does not contain data from fits to numerical Laplace inversions, as the large values of the hyperbolic functions in $\hat{\phi}(x,s)$ for $\epsilon=10^{-7}$ hinder its numerical Laplace inversion.
Both panels start showing the scaling $\tau_{\rm late}\propto 1/\sqrt{\delta}$ [\cref{eq:zhaoscaling}] around $\delta\approx10$. 
For $\epsilon=10^{-2}$ [\cref{fig:tau}(b)], however, \cref{eq:tau_late} also starts to deviate from the numerical results (blue dotted line) around that $\delta$.
After the $\tau_{\rm late}\propto 1/\sqrt{\delta}$ scaling, $\tau_{\rm late}$ reaches a minimum and then starts increasing, up to a  plateau around $4/\pi^2$ for extremely large $\delta$.
As anticipated, \cref{eq:tau_late} decreases monotonically with $\delta$, so it misses $\tau_{\rm late}$'s minimum around $\delta\approx20$ [in \cref{fig:tau}(b)] and subsequent increase.
Conversely, \cref{eq:tau_solution,eq:sol} excellently describe the numerical data up to $\delta\sim10^3$.
\Cref{eq:tau_solution,eq:sol} start to deviate from the approximation \cref{eq:tau_late} when the term $\mathcal{O}(\epsilon^2\delta)$ in \cref{eq:tau_late} can no-longer be neglected---that is, when the $\mathcal{O}(\epsilon^2\delta)$ term is as large as the first term of $\mathcal{O}(\epsilon\delta^{-1/2})$, which happens once $\delta\sim \epsilon^{-2/3}$.
Indeed, we start seeing discrepancies between the green and purple curves around $(10^{-2})^{-2/3}=21.5$ in \cref{fig:tau}(b) and around $(10^{-7})^{-2/3}=\SI{4.65e4}{}$ in \cref{fig:tau}(c).
Restoring units, we conclude that Zhao's scaling \cref{eq:zhaoscaling} \cite{zhao2011diffuse} does not hold generally for $\ell_c \gg\lambda_D$, but only as long as $\ell_c \ll L^{2/3}\lambda_D^{1/3}$.

After the local minimum, for $\delta\gg \epsilon^{-2/3}$, we observe that   $\tau_{\mathrm{late}}D/L^2$ increases proportional to $\delta$, as was predicted by \cref{eq:ourscaling}.

\subsubsection{Late-time response response for $\delta_1<\delta_2<\delta$}
Finally, when $\delta_2<\delta$, $s_{\mathrm{late}}$ lies on the interval $[s_{\epsilon},s_-]$.
\Cref{fig:realimag_gh}(c) shows that both $g$ and $h$ are purely imaginary there.
Expressing $h$ in terms of $g$ as $h=\sqrt{(g^2+\epsilon^{-2}-\gamma^{-2})/(\gamma^2g^2-1)}$, and writing $g=iG$ and $h=iH$, we find that  \cref{eq:pole} becomes
\begin{equation}\label{eq:ipole}
    H^3\tan G-G^3\tan H=GH\left[(GH\gamma\epsilon)^2-1\right](G^2-H^2)\,.
\end{equation}
For $\delta>\delta_2$, \cref{eq:ipole} has several solutions on $[s_{\epsilon},s_-]$.
Again, the late-time transient response is set by the solution $s_{\rm late}$ to \cref{eq:ipole} closest to $s=0$.
\Cref{eq:ipole} can only be solved numerically; we denote its solutions by $G_j$.
Then, the poles in $s$ can be expressed as
    $s_j=-(G_j^2+\epsilon^{-2}+\gamma^2G_j^4)/(\gamma^2G_j^2+1)$.
For $\delta\gg 1/\epsilon$, we find a solution $G\approx \pi/2$, corresponding to $s_{\mathrm{late}}=-\pi^2/4$.
Restoring units, we find $\tau_{\mathrm{late}}=4L^2/(\pi^2D)$.

The $\delta\gg 1/\epsilon$ regime can also be viewed as follows.
For $\delta\to \infty$, the potential $\phi(x,t)$ becomes linear, cf. \cref{fig:equilibrium}; hence, $\phi(x,t)=\Phi x$, independent of $t$.
\Cref{eq:contt} governing the ionic charge dynamics reduces to Fick's second law, whose characteristic timescale is $4L^2/\pi^2D$.
We seem to have arrived at a paradox, where we find the late-time relaxation time $4L^2/\pi^2D$ for the current relaxation (in \cref{fig:tau}), which directly derives from  $\phi(x,t)$, which is supposed to be time-independent.
This must be because the residue of the pole at $s=0$ dominates \cref{eq:Res}.
$s_{\mathrm{late}}=-\pi^2/4$ may be the location of the next-nearest pole coming from $s=0$, but for $\delta\to \infty$, that pole's contribution becomes negligible.

\subsection{Early-time relaxation}\label{sec:early}
 {The system's early-time response is set by the many poles on the negative $s$-axis, as previously mentioned, but not shown in }\cref{fig:polestructure}a.
 {However, we do not need to analyze these poles to determine the early-time response of $\sigma(t)$ and $\iota(t)$, for instance.
Instead, we use that $t\to 0$ corresponds to $s\to-\infty$.
For $s\to-\infty$, we find that $g= \gamma^{-1}+\mathcal{O}(s^{-1})$, $h= s^{1/2}+\mathcal{O}(s^{-1})$, and }\cref{eq:sigmalapl} { becomes}
\begin{figure}
	\centering
	\includegraphics[width = 0.98\linewidth]{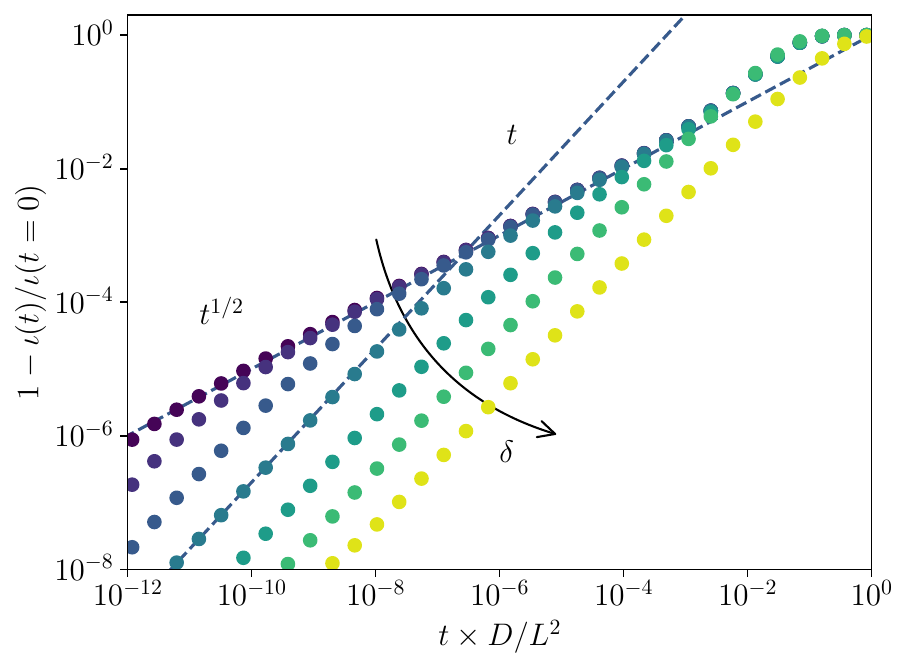}
    \caption{ {Transient electric current at early times for $\epsilon=1/20$ and $\delta={0,10^{-4},10^{-3},10^{-2},10^{-1},1,100}$ (from purple to yellow symbols). In contrast to }\cref{fig:transient_phi_and_I}b {, here we show ${1-\iota(t)/\iota(t=0)}$, to visualize early-time behavior better. The dashed lines are guides to the eye, indicating square-root and linear growth.}}\label{fig:early}
\end{figure}
\begin{align}\label{eq:sigmas}
    \hat{\sigma}(s)=c_1s^{-1}+c_2s^{-2}+c_3s^{-5/2}+c_4s^{-3}+\mathcal{O}(s^{-4}),
\end{align}
 {where $c_1,\dots,c_4$ are functions of $\gamma$, $\epsilon$, and $\delta$, but not of $s$.
Inverse Laplace transformation of }\cref{eq:sigmas} { yields}
\begin{align}
    \sigma(t)=c_1+c_2t+c_3t^{3/2}+c_4t^2+\mathcal{O}(t^3).
\end{align}
 {We obtain the current using $\iota(t)=\diff \sigma(t)/\diff t$,}
\begin{align}\label{eq:currearly}
    \iota(t)=c_2+c_3t^{1/2}+c_4t+\mathcal{O}(t^2).
\end{align}
\Cref{fig:early} { shows $1-\iota(t)/\iota(0)$ as obtained by numerical inverse Laplace transformation of }\cref{eq:sigmalapl} { for several $\delta$. Hence, this figure is similar to }\cref{fig:transient_phi_and_I}b,  {but by plotting $1-\iota(t)/\iota(0)$ instead of $\iota(t)$ we close in on $\iota(t)$'s temporal scaling (it eliminates $c_2$ in }\cref{eq:currearly}).
 {For strongly correlated systems (yellow symbols, $\delta=100$), $1-\iota(t)/\iota(0)\propto t$ until $t\sim L^2/D$, after which exponential decay sets in (cf. }\cref{fig:transient_phi_and_I}). 
 {For weakly correlated systems (greenish blue symbols, $\delta=0.1$), we observe $1-\iota(t)/\iota(0)\propto t$ at early times and $\propto t^{1/2}$ at intermediate times, in full agreement with }\cref{eq:currearly}.
 {In uncorrelated systems (purple symbols, $\delta=0$), the current scales as  $1-\iota(t)/\iota(0)\propto t^{1/2}$ until exponential relaxation sets in.} 
 {This square-root-in-time scaling of uncorrelated electrolytes also follows from Eqs.~(25) and (26) of Ref.}~\cite{bazant_pre_2004} { by the same asymptotic analysis as shown above.}
Ref.~\cite{janssen_pre_2018,palaia2019charged} { found exact expressions for the early-time response of uncorrelated systems, but as they expressed their results in infinite sums of exponentially decaying modes, they missed the system's early-time diffusive scaling.}

\section{Conclusions}
We have analyzed the response of a model electrical double layer capacitor subject to small a step potential difference.
Short-range electrostatic correlations are captured in our model through the BSK equation, containing a correlation length $\ell_c$.
Our main finding is that the late-time relaxation time $\tau_{\rm late}$ of the setup depends nonmonotonically on  $\ell_c$.
For $\ell_c\ll L^{2/3}\lambda_D^{1/3}$, we recover Zhao's prediction of $\tau_{\rm late}$ decreasing monotonically with increasing $\ell_c$ \cite{zhao2011diffuse}.
However, for  $\ell_c\sim L^{2/3}\lambda_D^{1/3}$ and beyond, $\tau_{\rm late}$ reaches a minimum, starts increasing, and reaches a plateau.
The plateau will be difficult to observe experimentally, as it happens for $\ell_c$ that are currently not accessible.
Conversely, the breakdown of Zhao's scaling \cref{eq:zhaoscaling} and the subsequent $\tau_{\rm late}$ minimum around $\ell_c\sim L^{2/3}\lambda_D^{1/3}$ should be accessible for highly correlated, highly confined electrolytes, for instance, ionic liquids in the surface force balance apparatus.
While our work focused on the ramifications of the BSK equation, future work could study the relaxation times of other recent correlated-electrolyte models \cite{desouza_jpcc_2020,souza2020interfacial,gupta_jpcc_2020,gupta_prl_2020}.

\section{Acknowledgements}
We thank Jeffrey Everts and Svyatoslav Kondrat for inspiring discussions.
Both authors were supported by a FRIPRO grant from The Research Council of Norway (Project No. 345079).\\

\section{Data availability}
 {All routines used in the creation of the figures are available at }\href{https://github.com/fdavid92/bsk-charging/tree/main}{https://github.com/fdavid92/bsk-charging/tree/main}

\appendix
\renewcommand\thefigure{\thesection\arabic{figure}} \setcounter{figure}{0}  

\section{Late-time relaxation with boundary conditions of Ref.~\cite{desouza_jpcc_2020}}\label{app:A}

 {De Souza and Bazant (SB) derived the boundary condition $\delta\partial_x^3\phi=\pm\partial_x^2\phi\big|_{x=\pm 1}$, based on mechanical equilibrium at the electrode-electrolyte interface, replacing the BSK boundary condition $\delta\partial_x^3\phi\Big|_{x=\pm1}=0$} [\cref{eq:bc3rdorder}], { which lacked a physical argument} \cite{desouza_jpcc_2020}.
 {The limit $\delta\to 0$ seems to imply $\partial_x^2\phi=0$, which does not appear as a boundary condition for the Debye-Falkenhagen equation} \cite{bazant_pre_2004,janssen_pre_2018}. 
 {However, assuming mechanical equilibrium at the interface does not yield additional boundary conditions when $\delta=0$, rendering analyzing any upcoming solution with the SB boundary condition meaningless for $\delta=0$.}

 {Solving }\cref{Eq:4Poiss} { with the SB boundary condition yields}
\begin{align}
    \hat{\phi}=&\dfrac{\Phi}{s\Xi}\bigg\{\dfrac{\sinh gx}{\cosh g }-gx \left[\delta^2\varepsilon^4g^4-\dfrac{\varepsilon^2}{\delta}g\tanh g+1\right.\nn
    &\left.-\dfrac{g}{h}\dfrac{\delta g - \tanh g}{\delta h - \tanh h}\left(\delta^2\varepsilon^4h^4-\dfrac{\varepsilon^2}{\delta}h\tanh h+1\right)\right]\nn
    &-\dfrac{g^2 \sinh hx}{h^2 \cosh h}\dfrac{\delta g - \tanh g}{\delta h - \tanh h}\bigg\}\nn
    \intertext{with}\nn
    \Xi =& \tanh g - g\left[\delta^2\varepsilon^4g^4-\dfrac{\varepsilon^2}{\delta}g\tanh g+1-\dfrac{g}{h}\dfrac{\delta g - \tanh g}{\delta h - \tanh h}\right. \nn
    &\left.  \times\left(\delta^2\varepsilon^4h^4-\dfrac{\varepsilon^2}{\delta}h\tanh h+1-\dfrac{\tanh h}{h}\right)\right].
\end{align}
 {As before, we determine the surface charge density $\sigma$ from $\hat{\phi}$. Note, however, that the derivation of }\cref{eq:sigmaP} { involved using $\partial_x^3\hat{\phi}\big|_{x=\pm1}=0$. For the SB boundary condition, we find}
\begin{align}\label{eq:BSKdSsig}
\hat{\sigma}\Big|_{x=\pm1}=\pm\epsilon^2(\partial_x\hat{\phi}+\delta^2\partial_x^3\hat{\phi})
\end{align}
 {instead.}
\begin{figure}[b]
	\centering
	\includegraphics[width = \linewidth]{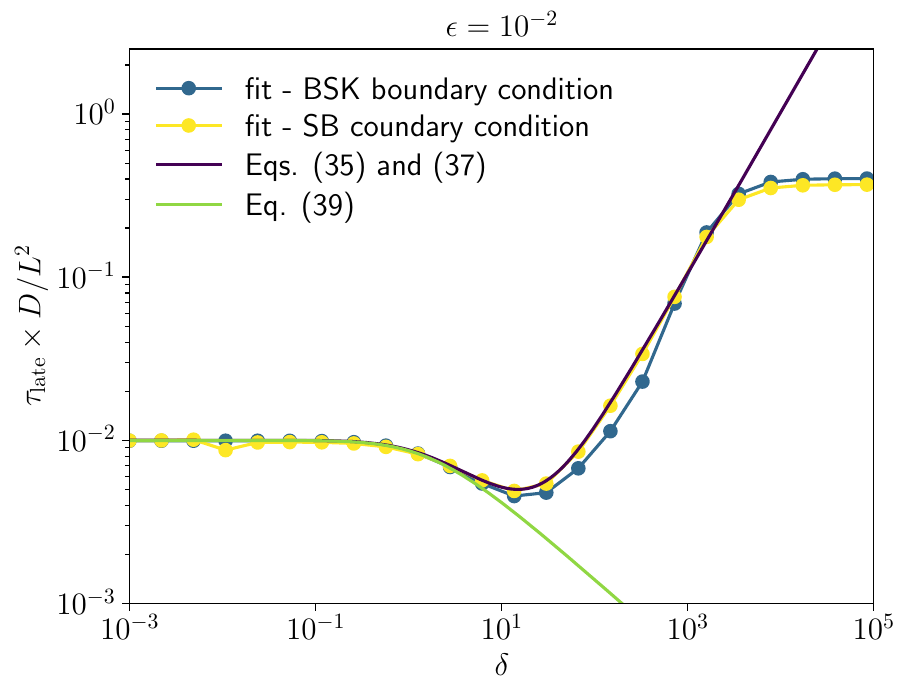}
    \caption{ {Relaxation time as determined by fitting the exponential decay of numerically obtained $\iota(t)$ using the BSK boundary conditions (blue) and the SB boundary conditions (yellow). Also shown are $\tau_{\mathrm{late}}$ predictions from }\cref{eq:tau_solution,eq:sol} { (purple) and }\cref{eq:tau_late} { (green) for  $\epsilon=10^{-2}$.}}\label{fig:dSBC}
\end{figure}

\Cref{fig:dSBC}  {shows $\tau_{\mathrm{late}}$ as obtained using numerical inverse Laplace transform of the areal electronic current $\hat{\iota}(s)=s\hat{\sigma}(s)$ for the BSK and SB boundary conditions and the analytical approximations of} \cref{eq:tau_solution,eq:sol,eq:tau_late}.
 {We see that the time constant resulting from the SB boundary conditions agrees quantitatively with the one from BSK theory. 
Hence, the conclusions drawn in the main body of the paper are unchanged.}


\begin{thebibliography}{10}

\bibitem{debye_pz1_1923}
P.~Debye and E.~Hückel.
\newblock Zur {Theorie} der {Elektrolyte}. {I.} {Gefrierpunktserniederung} und
  verwandte {Erscheinungen}.
\newblock {\em Phys. Z.}, 24:185--206, 1923.

\bibitem{levin2002electrostatic}
Y.~Levin.
\newblock Electrostatic correlations: From plasma to biology.
\newblock {\em Rep. Prog. Phys.}, 65(11):1577, 2002.

\bibitem{santos_ea_2024}
E.~Santos and W.~Schmickler.
\newblock On the timescale of electrochemical processes.
\newblock {\em Electrochim. Acta}, 498:144659, 2024.

\bibitem{debye_pz_1923}
P.~Debye and E.~Hückel.
\newblock Zur {Theorie} der {Elektrolyte}. {II.} {Das} {Grenzgesetz} für die
  elektrische {Leitfähigkeit}.
\newblock {\em Phys. Z.}, 24:305--325, 1923.

\bibitem{onsager_pz_1926}
L.~Onsager.
\newblock Zur {Theorie} der {Elektrolyte} {I}.
\newblock {\em Phys. Z.}, 27:388--392, 1926.

\bibitem{onsager_pz_1927}
L.~Onsager.
\newblock Zur {Theorie} der {Elektrolyte} {II}.
\newblock {\em Phys. Z.}, 28:277--298, 1927.

\bibitem{debye_zeapc_1928}
P.~Debye and H.~Falkenhagen.
\newblock Dispersion der {Leitfähigkeit} starker {Elektrolyte}.
\newblock {\em Z. Elektrochem. angew. phys. Chem}, 34(9):562--565, 1928.

\bibitem{macdonald_pr_1953}
J.~R. Macdonald.
\newblock Theory of ac space-charge polarization effects in photoconductors,
  semiconductors, and electrolytes.
\newblock {\em Phys. Rev.}, 92(1):4–17, 1953.

\bibitem{bazant_pre_2004}
M.~Z. Bazant, K.~Thornton, and A.~Ajdari.
\newblock {Diffuse-charge} dynamics in electrochemical systems.
\newblock {\em Phys. Rev. E}, 70:021506, 2004.

\bibitem{janssen_pre_2018}
M.~Janssen and M.~Bier.
\newblock Transient dynamics of electric double-layer capacitors: {E}xact
  expressions within the {Debye-Falkenhagen} approximation.
\newblock {\em Phys. Rev. E}, 97:052616, 2018.

\bibitem{palaia2019charged}
I.~Palaia.
\newblock {\em Charged systems in, out of, and driven to equilibrium: from
  nanocapacitors to cement}.
\newblock PhD thesis, Universit{\'e} Paris Saclay (COmUE), 2019.

\bibitem{asta_jcp_2019}
A.~J. Asta, I.~Palaia, E.~Trizac, M.~Levesque, and B.~Rotenberg.
\newblock Lattice {Boltzmann} electrokinetics simulation of nanocapacitors.
\newblock {\em J. Chem. Phys.}, 151(11):114104, 2019.

\bibitem{scalfi_arpc_2021}
L.~Scalfi, M.~Salanne, and B.~Rotenberg.
\newblock Molecular simulation of electrode-solution interfaces.
\newblock {\em Annu. Rev. Phys. Chem.}, 72:189--212, 2021.

\bibitem{ahrensivers_jcp_2022}
L.~J.~V. Ahrens-Iwers, M.~Janssen, S.~R. Tee, and R.~H. Meißner.
\newblock {ELECTRODE:} {An} electrochemistry package for atomistic simulations.
\newblock {\em J. Chem. Phys.}, 157(8):084801, 08 2022.

\bibitem{pireddu_prl_2023}
G.~Pireddu and B.~Rotenberg.
\newblock Frequency-dependent impedance of nanocapacitors from electrode charge
  fluctuations as a probe of electrolyte dynamics.
\newblock {\em Phys. Rev. Lett.}, 130:098001, 2023.

\bibitem{beunis_apl_2007}
F.~Beunis, F.~Strubbe, M.~Marescaux, K.~Neyts, and A.~R.~M. Verschueren.
\newblock Diffuse double layer charging in nonpolar liquids.
\newblock {\em Appl. Phys. Lett.}, 91(18):182911, 2007.

\bibitem{zhao_jpcc_2024}
C.~Zhao, T.~Yang, S.~Jin, and B.~Wu.
\newblock Measurement of electric double layer charging dynamics on platinum
  electrodes in aqueous solutions of alkali sulfates and nitrates.
\newblock {\em J. Phys. Chem. C}, 128(14):5964--5971, 2024.

\bibitem{kortschot_jpcc_2014}
R.~J. Kortschot, A.~P. Philipse, and B.~H. Ern{\'e}.
\newblock Debye length dependence of the anomalous dynamics of ionic double
  layers in a parallel plate capacitor.
\newblock {\em J. Phys. Chem. C}, 118(22):11584--11592, 2014.

\bibitem{nakamura2014structural}
M.~Nakamura, H.~Kaminaga, O.~Endo, H.~Tajiri, O.~Sakata, and N.~Hoshi.
\newblock Structural dynamics of the electrical double layer during capacitive
  charging/discharging processes.
\newblock {\em J. Phys. Chem. C}, 118(38):22136--22140, 2014.

\bibitem{ojha2020double}
K.~Ojha, N.~Arulmozhi, D.~Aranzales, and M~T.~M. Koper.
\newblock Double layer at the {Pt}(111)–aqueous electrolyte interface:
  Potential of zero charge and anomalous {Gouy–Chapman} screening.
\newblock {\em Angew. Chem. - Int. Ed.}, 59(2):711--715, 2020.

\bibitem{fertig_pccp_2020}
D.~Fertig, M.~Valisk\'{o}, and D.~Boda.
\newblock Rectification of bipolar nanopores in multivalent electrolytes:
  {E}ffect of charge inversion and strong ionic correlations.
\newblock {\em Phys. Chem. Chem. Phys.}, 22:19033--19045, 2020.

\bibitem{bazant_pre_2011}
M.~Z. Bazant, B.~D. Storey, and A.~A. Kornyshev.
\newblock Double layer in ionic liquids: {O}verscreening versus crowding.
\newblock {\em Phys. Rev. Lett.}, 106:046102, 2011.

\bibitem{storey_pre_2012}
B.~D. Storey and M.~Z. Bazant.
\newblock Effects of electrostatic correlations on electrokinetic phenomena.
\newblock {\em Phys. Rev. E}, 86:056303, 2012.

\bibitem{Hatlo_2012}
M.~M. Hatlo, R.~van Roij, and L.~Lue.
\newblock The electric double layer at high surface potentials: The influence
  of excess ion polarizability.
\newblock {\em Europhys. Lett.}, 97(2):28010, 2012.

\bibitem{balu2018role}
B.~Balu and A.~S. Khair.
\newblock Role of {Stefan–Maxwell} fluxes in the dynamics of concentrated
  electrolytes.
\newblock {\em Soft Matter}, 14:8267--8275, 2018.

\bibitem{desouza_jpcc_2020}
J.~P. de~Souza and M.~Z. Bazant.
\newblock Continuum theory of electrostatic correlations at charged surfaces.
\newblock {\em J. Phys. Chem. C}, 124(21):11414--11421, 2020.

\bibitem{souza2020interfacial}
J.~P. de~Souza, Z.~A.~H. Goodwin, M.~McEldrew, A.~A. Kornyshev, and M.~Z.
  Bazant.
\newblock Interfacial layering in the electric double layer of ionic liquids.
\newblock {\em Phys. Rev. Lett.}, 125:116001, 2020.

\bibitem{gupta_jpcc_2020}
A.~Gupta, A.~R. Govind, E.~A. Carter, and H.~A. Stone.
\newblock Thermodynamics of electrical double layers with electrostatic
  correlations.
\newblock {\em J. Phys. Chem. C}, 124(49):26830--26842, 2020.

\bibitem{gupta_prl_2020}
A.~Gupta, A.~R. Govind, E.~A. Carter, and H.~A. Stone.
\newblock Ionic layering and overcharging in electrical double layers in a
  {Poisson-Boltzmann} model.
\newblock {\em Phys. Rev. Lett.}, 125:188004, 2020.

\bibitem{zhao2011diffuse}
H.~Zhao.
\newblock Diffuse-charge dynamics of ionic liquids in electrochemical systems.
\newblock {\em Phys. Rev. E}, 84:051504, 2011.

\bibitem{alijo2015effects}
P.~H.~R. Alijó, F.~W. Tavares, E.~C. {Biscaia Jr.}, and A.~R. Secchi.
\newblock Effects of electrostatic correlations on ion dynamics in alternating
  current voltages.
\newblock {\em Electrochim. Acta}, 152:84--92, 2015.

\bibitem{lee_prl_2015}
A.~A. Lee, S.~Kondrat, D.~Vella, and A.~Goriely.
\newblock Dynamics of ion transport in ionic liquids.
\newblock {\em Phys. Rev. Lett.}, 115:106101, 2015.

\bibitem{yochelis_pccp_2020}
A.~Yochelis.
\newblock Transition from non-monotonic to monotonic electrical diffuse layers:
  impact of confinement on ionic liquids.
\newblock {\em Phys. Chem. Chem. Phys.}, 16:2836--2841, 2014.

\bibitem{ma_jcp_2022}
K.~Ma, M.~Janssen, C.~Lian, and R.~van Roij.
\newblock Dynamic density functional theory for the charging of electric double
  layer capacitors.
\newblock {\em J. Chem. Phys.}, 156(8):084101, 2022.

\bibitem{Note1}
We added the length $\ell _c$ in \protect \cref {eq:bc3rdorder} to ensure that,
  in the limit of uncorrelated electrolytes $\ell _c\to 0$, our system of
  \protect \cref {eq:base,eq:initialcondition,eq:boundaryconditions} reduces to
  those of Refs.~\cite {bazant_pre_2004,janssen_pre_2018}.

\bibitem{valisko_aip_2018}
M.~Valisk\'{o}, T.~Krist\'{o}f, D.~Gillespie, and D.~Boda.
\newblock A systematic {Monte Carlo} simulation study of the primitive model
  planar electrical double layer over an extended range of concentrations,
  electrode charges, cation diameters and valences.
\newblock {\em AIP Adv.}, 8(2):025320, 2018.

\bibitem{Note2}
We determined $s_+,s_-$, and $s_{\epsilon }$ analytically and $s_{\protect
  \mathrm {late}}$ numerically.

\bibitem{Note3}
For $\delta =0$, our model reduces to the one in Ref.~\cite {janssen_pre_2018},
  and the poles are identical to those of that paper.

\bibitem{Note4}
For $\gamma \rightarrow 0$, the left-hand side of \protect \cref
  {eq:needtosolve} tends to 1, reducing to an equation at the bottom of page 4
  of Ref.~\cite {janssen_pre_2018}.

\end{thebibliography}
\bibliographystyle{unsrt}

\end{document}